\documentclass[aps,twocolumn,showpacs,floatfix]{revtex4}

\usepackage{graphicx}

\begin{document}
\title{First principles modeling of tunnel magnetoresistance of Fe/MgO/Fe 
trilayers}

\author{Derek Waldron$^1$, Vladimir Timoshevskii$^1$, Yibin Hu$^{1,2}$, 
Ke Xia$^{2,1}$ and Hong Guo$^1$}
\affiliation{
1. Centre for the Physics of Materials and Department of Physics, McGill
University, Montreal, PQ, H3A 2T8, Canada\\
2. International Center for Quantum Structures (ICQS), Institute of
Physics, Chinese Academy of Science, Beijing, China}

\begin{abstract}

By carrying out density functional theory analysis within the Keldysh 
non-equilibrium Green's functional formalism, we have calculated the nonlinear 
and non-equilibrium quantum transport properties of Fe/MgO/Fe trilayer 
structures as a function of external bias voltage. For well relaxed atomic 
structures of the trilayer, the equilibrium tunnel magnetoresistance ratio 
(TMR) is found to be very large and also fairly stable against small 
variations in the atomic structure.  As a function of external bias voltage, 
the TMR reduces monotonically to zero with a voltage scale of about 1V, in 
agreement with experimental observations. We present understanding of the 
nonequilibrium transport properties by investigating microscopic details of the 
scattering states and the Bloch bands of the Fe leads. 

\end{abstract}

 \maketitle

Since the prediction and elegant physics explanation\cite{zhang1,mathon} 
that magnetic tunnel junction (MTJ) of Fe/MgO/Fe trilayer structure may have 
very high tunnel magnetoresistance (TMR), MgO based MTJ has progressed 
at a rapid pace in recent years and produced the highest measured TMR at room 
temperature: several groups\cite{parkin1,yuasa} reported TMR ratio in the 
range of 180\% to 250\%. TMR effect presents an excellent opportunity for 
spintronics, it is the key to magnetoresistive 
random-access-memory\cite{moodera2}, programmable logic elements\cite{ploog}, 
and magnetic sensors. The high TMR values in MgO based MTJs have generated 
great excitement for practical applications.

Atomistic calculations\cite{zhang1,mathon} have so far played a vital role
in elucidating the reason behind the observed large TMR in Fe/MgO/Fe 
MTJs\cite{zhang1,mathon,parkin1,yuasa}. There are, however, a number of 
important issues remain to be understood from atomic first principles.
Most existing work predicted\cite{zhang1,mathon} TMR to be greater than 
1000\%, experimental data are still lower. More seriously is that experimental 
data on MgO based MTJs show a monotonically \emph{decreasing} TMR as a 
function of applied bias voltage\cite{yuasa,parkin1} and it reduces to zero 
when bias is about one volt. To the best of our knowledge, there have been 
two atomistic calculations of bias dependence of TMR for MgO 
barriers\cite{zhang2,heiliger}, both used the Korringa-Kohn-Rostoker
numerical technique. Ref.\onlinecite{zhang2} predicted a substantial
\emph{increase} of TMR versus bias for the asymmetric system analyzed there,
while Ref.\onlinecite{heiliger} found a roughly constant TMR, a decaying 
TMR, or an entirely negative TMR versus bias depending on atomic structures 
of the interface. The origin of these differences were not clear. Earlier
theory\cite{levy} on $Al_2O_3$ based MTJs has attributed small bias
dependence of magneto-resistance to magnon scattering. Given the extreme 
importance of MgO based MTJ in near future spintronics and the accumulated 
experimental data, further \emph{quantitative} understanding on quantum 
transport in Fe/MgO/Fe at finite bias is urgently needed. 

Here we present a first principles atomistic analysis of nonlinear and
non-equilibrium quantum transport in Fe/MgO/Fe MTJ. We use a state-of-the-art 
quantum transport technique\cite{mcdcal,matdcal} which is based on real-space, 
Keldysh nonequilibrium Green's function (NEGF) formalism combined with density 
functional theory (DFT). The basic idea of the NEGF-DFT formalism\cite{mcdcal} 
is to calculate device Hamiltonian and electronic structure by DFT, populate 
this electronic structure using NEGF which properly takes into account 
nonequilibrium quantum statistics, and deal with open device boundaries 
directly using real-space numerical techniques. The power of NEGF-DFT methods 
have already been demonstrated by direct quantitative comparison to 
experimental data\cite{kaun1}. Our results show that for fully relaxed 
atomic structure of the Fe/MgO/Fe device, the equilibrium TMR ratio reaches 
several thousand percent---consistent with previous theoretical 
results\cite{zhang1,kirschner}. This value is also found to be fairly stable 
against small variations in atomic structure. We found that the TMR ratio 
is monotonically quenched by the bias $V_b$ with a scale of about one volt.
The microscopic details of these transport features can be understood by the 
behavior of bias dependent scattering states.

\begin{figure}
\centering
\includegraphics[width=1.0\linewidth]{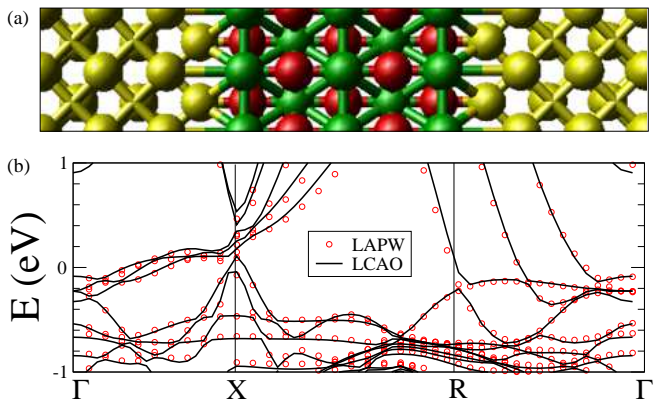}
\caption{ (color online)
(a) Schematic plot of a two probe Fe(100)/MgO(100)/Fe(100) device. The system 
has infinite extent in the (x,y) direction with a lattice constant of 
2.82\AA, and extends to $\pm \infty$ in the z-direction. (b) Band structure 
of a periodic $\cdots$Fe/MgO/Fe/MgO$\cdots$ lattice obtained using 
optimized LCAO pseudopotentials and basis sets compared to that from 
full potential LAPW method. A good agreement is found to be necessary 
in order to carry out the NEGF-DFT analysis for the two probe Fe/MgO/Fe devices.
}
\label{fig1}
\end{figure}

The MTJ is schematically shown in Fig.1a where a number of MgO(100) layers is 
sandwiched by two Fe(100) leads. The MTJ is periodic in the $x-y$ direction 
while the leads also extend to $z=\pm \infty$ (transport direction). For DFT, 
we use standard norm-conserving pseudopotentials\cite{pseudopotentials} and an 
{\it s, p, d} double-zeta LCAO basis set\cite{siesta}. The exchange-correlation 
potential is treated at the LSDA level. In order to accurately determine 
transport properties of the MTJs, we found that special care must be given to 
the pseudopotentials and basis sets. In our work, these inputs were carefully 
constructed to accurately reproduce the electronic structures of Fe, MgO, 
{\it and} periodic lattice of the Fe/MgO interface obtained by full potential 
LAPW method\cite{LAPW}. The latter comparison is shown in Fig.1b.

For the two probe MTJ simulation, we found that $20 \times 20$ ($k_x,k_y$) 
points suffice to sample the 2D transverse Brillouin zone (BZ) for converging 
the density matrix on the {\it complex contour} energy integration 
in the NEGF-DFT self-consistent analysis\cite{mcdcal}. Much denser k-sampling 
of $10^6$ ($k_x,k_y$) points was required to converge the density matrix for 
the real energy integration of NEGF\cite{mcdcal,matdcal} and for computing 
the transmission coefficient by summing over the BZ: $T_{\sigma}(E,V_b)\ = \
\sum_{k_x,k_y} T_{\sigma}^{k_x,k_y}(E,V_b)$. Here $E$ is the electron energy,
$V_b$ the external bias. The BZ resolved transmission, $T_{\sigma}^{k_x,k_y}$, 
is obtained by standard Green's functions technique: $T_{\sigma}^{k_x,k_y}
\equiv Tr [ Im({\bf \Sigma}_L^r){\bf G}^r Im({\bf \Sigma}_R^r){\bf G}^a ]$,
where all quantities in the trace are functions of transverse momentum. Here
$\sigma\equiv \uparrow,\downarrow$ is the spin index; ${\bf G}^{r,a}$ are the
retarded/advanced Green's function matrices in spin and orbital space; and
${\bf \Sigma}_{L,R}^r$ are the retarded self-energies due to the existence
of the bulk-3D left/right Fe leads. Finally, the spin-current (spin polarized 
charge current) is obtained by $I_\sigma (V_b)=\frac{e}{h}\int_{\mu_L}^{\mu_R} 
dE T_{\sigma}(E,V_b)[ f_L(E-\mu_L) -f_R(E-\mu_R)]$ where $\mu_{L,R}$ is the 
electrochemical potential of the left/right leads. The total charge current 
is given by $I\equiv I_{\uparrow}+I_{\downarrow}$. In our calculations, the 
atomic structure was fully relaxed by the LAPW\cite{LAPW} method between three 
Fe layers on each side of the MgO, with the most remote layer of Fe atoms 
fixed at crystalline positions during relaxation. The x-y lattice constant 
$a$ of the interface was fixed to our {\it calculated} one for bcc Fe, 
$a=2.82$\AA. The Fe-O distance was found to be $2.236$\AA for a completely 
relaxed structure in agreement with previous studies\cite{zhang1}.

Fig.2a,b plots the current-voltage (I-V) characteristics (solid line) for 
5-layer MgO device in the parallel magnetization configuration (PC) and the 
anti-parallel configuration (APC) of the two Fe leads, respectively. The lower
insets plot the majority spin-current at small bias range. For bias 
less than $0.8V$, the total current remains extremely small. At about $1.5V$, 
the device ``turns on'' and current increases rapidly afterward. Such a 
turn-on voltage is consistent with experimental data\cite{MgO1,yuasa}. The 
spin-currents are shown as the dashed and dotted lines for the up- and 
down-channels (majority-, minority-channel). We found that the initial rise 
of the current at $\sim 0.8V$ in PC is dominated by the down-channel where 
$I_{\downarrow}$ exceeds $I_{\uparrow}$ by over a factor of eight. This can be 
explained by investigating the transmission coefficients (see below). Above 
$\sim 1.5V$ the spin-currents roughly contribute equally to the total current. 
The I-V curves for a 3-layer MgO device are plotted in the inset of Fig.2a,b 
and show similar features.

\begin{figure}
\centering
\includegraphics[width=1.0\linewidth]{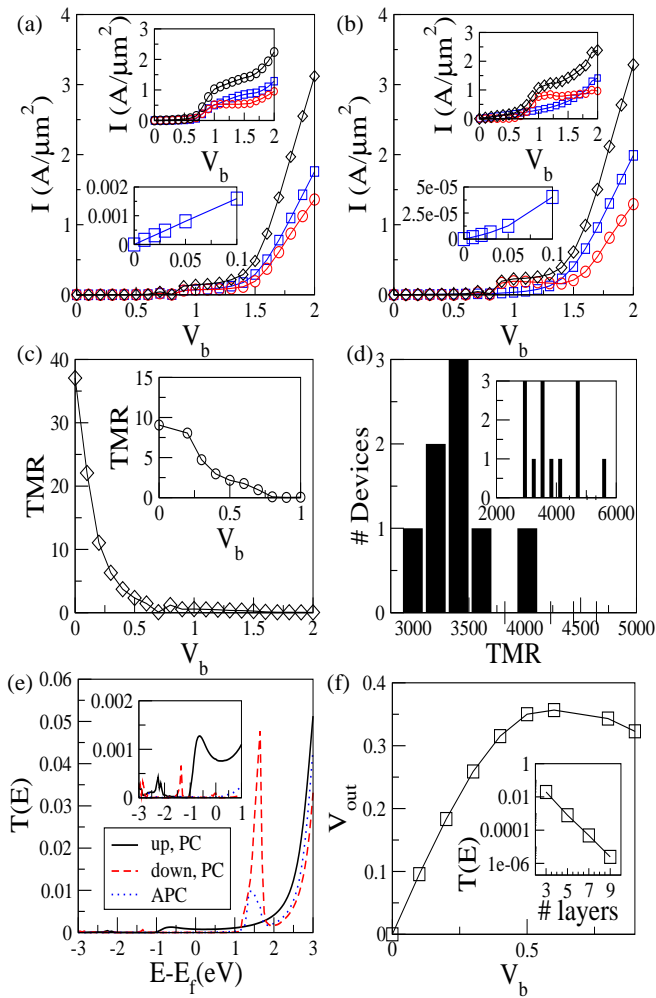}
\caption{ (color online)
(a,b) I-V curves for 5-layer PC and APC, respectively. Solid line (diamonds): 
total current; dashed line (squares): $I_\uparrow$; dotted line
(circles): $I_\downarrow$. Upper inset: I-V curves for a 3-layer device. Lower 
inset: majority current $I_\uparrow$ vs $V_b$ for small ranges of $V_b$. 
(c) TMR vs bias $V_b$ for a 5-layer device. Inset: TMR for a 3-layer device. 
(d) Histogram of TMR for several 5-layer devices with variation in 
the position of the surface atoms. Inset: Histogram for varying all the Mg 
and O atoms in the device.
(e) Transmission coefficient $T_\sigma$ versus energy $E$ for $V_b=0$.
$E=0$ is the Fermi energy of leads. Solid: $T_\uparrow$ for PC setup;
dashed: $T_\downarrow$ for PC; dotted: $T_\uparrow=T_\downarrow$ for APC.
Inset: the same transmission coefficients at energies between $-3$eV and $1$eV.
(f) The magnitude of the output signal modulation $V_{out}$ as a function 
of bias. Inset: semi-log plot of zero bias total transmission coefficients 
$T$ versus the number of MgO layers for PC setup, indicating an exponential 
decrease of $T$. 
}
\label{fig2}
\end{figure}

From the I-V curves we infer a TMR ratio using the common {\em optimistic}  
definition: $R_{TMR}\equiv (I_{APC}-I_{PC})/I_{APC}$, where $I_{APC,PC}$ 
are the total currents in APC and PC respectively. At $V_b=0$ when all currents
vanish, we compute $R_{TMR}$ use transmission coefficients\cite{blugel}. From 
Fig.2c, for 5-layer MgO device $R_{TMR}\sim 3700$\% at zero bias and declines 
quickly with $V_b$, essentially vanishes on a scale of about $1V$. For the 
3-layer MgO we found $R_{TMR}\sim 850$\% at zero bias and declines to zero with 
$V_b$ on a similar bias scale. While the experimentally measured TMR has
increased dramatically in the past two years\cite{parkin1,yuasa}, they are
still significantly lower than theoretically predicted values here and 
elsewhere\cite{zhang1,mathon}. It is anticipated that surface roughness is 
playing a major role\cite{xia}. To investigate this effect, we generated 
eight device atomic structures of 5-layer MgO: for each device 
we varied the z-coordinates of the \emph{surface} Mg and O atoms from their
relaxed positions, by a random displacement corresponding to about 1\% of 
the bond length. Self-consistent NEGF-DFT analysis is carried out for 
them and the result is shown in Fig.2d. Of these eight atomic structures, 
the minimum TMR is about 3000\% while the maximum is $\sim 4000$\%, with 
an average of 3580\%. Although the sample size is small, the TMR ratio 
appears rather stable against small interface atom displacements. A 
similar analysis is carried out for thirteen 5-layer MgO devices where 
all the Mg and O atoms were displaced randomly by roughly 1\% of 
the bond length, the result is in the inset of Fig.2d. Again, even though 
the sample size is small, the results nevertheless indicate that small 
random variations of atomic positions in the barrier layer are not 
sufficient to reduce the zero bias TMR to the presently measured experimental 
values. Other mechanisms such as oxidization of the Fe surface, diffusive 
impurities and/or defects, are likely responsible for experimental 
TMR values. However our results provide a theoretical upper limit which 
does suggest that if a device can be manufactured with a high quality 
interface, it may be possible to achieve even higher TMR values than
presently known\cite{yuasa}. 

We now investigate nonequilibrium features, namely the bias dependence of
various transport properties of the MTJ. Fig.2c shows a dramatic 
quench of TMR by the external bias voltage with a scale of about one volt, 
in agreement with the experimental data\cite{yuasa,tiusan}. The origin of the
TMR quenching is due to a very fast rise in the APC current relative to 
the PC current as a function of bias. We now analyze these features. 

First, the voltage dependence of the total current and spin-current (Fig.2a,b) 
can be understood from the behavior of the transmission coefficient $T_\sigma$.
Fig.2e plots $T_{\sigma}=T_\sigma(E)$ versus electron energy $E$ at zero 
bias for PC and APC of the 5-layer MgO device. In PC, the majority carrier 
transmission $T_\uparrow$ (solid line) is smooth and several orders of 
magnitude larger than $T_\downarrow$ (dashed line) when $E$ is near the Fermi 
energy of the leads ($E_f=0$). By analyzing the spin-dependent scattering 
states\cite{mcdcal} of the MTJ, we were able to determine which bands of the 
Fe leads contribute to the transmission. We found that $T_\uparrow$ is 
dominated by the $\Delta_1$ band of the Fe leads, in agreement with 
Ref.\onlinecite{zhang1}. Below $-1eV$, $T_\uparrow$ becomes extremely small 
due to the disappearance of the $\Delta_1$ band. The $T_\downarrow$, on the 
other hand, is considerably less smooth because the transmission near the 
Fermi level is mostly dominated by interface resonance states\cite{Wunnicke}. 
In particular, a large peak in $T_\downarrow$ appears above $E=1eV$: as $E$ 
is increased, different Fe bands may participate transport and this peak is 
due to such a contribution. This $T_\downarrow$ peak explains the much 
larger minority-channel current than the majority-channel current in PC at 
$V_b=0.8V$ (Fig.2a), as already noted above. 

Second, for APC, we obtain $T_{\uparrow}=T_{\downarrow}$ for all $E$ at zero 
bias due to the geometrical symmetry of the device (dotted line in Fig.2e). 
We found that the BZ resolved total transmission, $T^{k_x,k_y}(E,V_b)=
T_\uparrow^{k_x,k_y}+T_\downarrow^{k_x,k_y}$ shown in Fig.3c for $V_b=0$ and 
Fig.3d for $V_b=0.05$V, is dominated by broad and smooth peaks at around 
$|k_x|=|k_y|=0.12$ (in unit of $\pi/a$ throughout the paper, where $a$ is the 
Fe lattice constant mentioned above), and there is almost no transmission at 
$k_x=k_y=0$. For $V_b=0$, Fig.3c also shows that the dominating peaks are 
surrounded by other much sharper peaks. For APC, it is the majority channel 
from one Fe layer traversing the MgO barrier and going to the minority channel 
in the other Fe layer. Figs.3a,b plot the majority and minority electronic 
band structures of Fe near $E_f$ for $|k_x|=|k_y|=0.12$, respectively. 
By projecting scattering states with $|k_x|=|k_y|=0.12$ onto the Fe bands of 
Fig.3a,b, the dominating peaks are found to be largely due to {\it channel} 
transmission: they are due to the band labelled ``1" in Fig.3a at one Fe 
contact, transmitting to the band labelled ``2" on the other Fe contact. 
Our calculations show that this band-to-band transmission 
contributes $2.37\times 10^{-4}$ to majority channel $T_\uparrow^{k_x,k_y}$. 
Other band-to-band transmissions are considerably smaller. Similarly, 
$T_\downarrow^{k_x,k_y}$ is mainly contributed from band-2 to band-1. 
Therefore, it is the band-to-band transmissions which give almost the entire 
hight of the dominating peaks in Fig.3c (note Fig.3c,d plot the total
BZ resolved transmission)\cite{foot2}. 

Third, we found that bias voltage has dramatic effects for APC. The very sharp 
peaks in Fig.3c, which are due to interface resonances occurring at zero bias,
are completely removed by a finite bias of 0.05V, as shown in Fig.3d. Moreover,
the dominating peaks become considerably higher than those in Fig.3c. We
checked that even 0.01V bias can remove these sharp resonances. Again, we 
found that transmission from band-1 of left Fe lead to band-2 of right lead 
dominates$T_\uparrow^{k_x,k_y}$, contributing $3.2\times 10^{-4}$ to the
peaks. This value is considerably larger than the value at zero bias,
indicating that bias enhances the coupling of Fe bands across the MgO barrier 
for APC, causing a fast relative increase in the APC current as a function of 
bias voltage, and is responsible for quenching the TMR observed in Fig.2c. 

\begin{figure}
\centering
\includegraphics[width=1.0\linewidth]{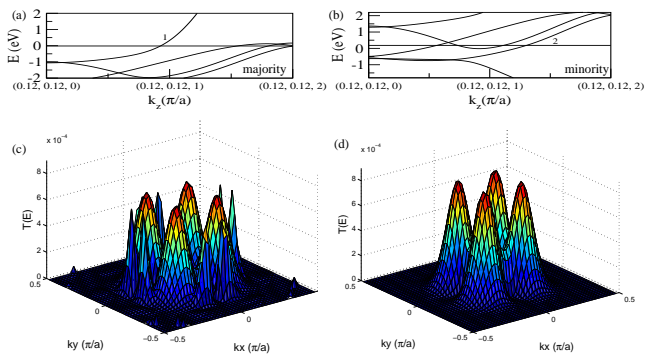}
\caption{ (color online)
(a,b) Fe bands at $|k_x|=|k_y|=0.12$ versus $k_z$ for majority and
minority electrons, respectively. (c,d) Total BZ resolved transmission 
coefficient at $E_f$ versus $k_x,k_y$, for 5-layer MgO. (c) for $V_b=0$; (d) 
for $V_b=0.05$V. The dominant peaks are near $|k_x|=|k_y|=0.12$. 
}
\label{fig3}
\end{figure}

An important parameter which has been measured\cite{yuasa,tiusan} is the 
magnitude of the output signal modulation, namely 
$V_{out}\equiv V_b(R_{APC}-R_{PC})/R_{APC}$ with $V_b$ the applied bias, 
$R_{APC}$ and $R_{PC}$ are the junction resistances for APC and PC. $V_{out}$ 
puts more weight on information of TMR for higher bias, and is plotted 
as a function of bias in Fig.2f. We found that $V_{out}$ increases in a roughly 
linear manner and then bends over at around $V_b\sim 0.5-0.7V$ where $V_{out}$ 
is about $350mV$. These voltage scales are rather similar to the experimental 
data\cite{yuasa,tiusan}. Finally, the inset of Fig.2f shows a semi-log plot 
of the zero bias total transmission at Fermi energy versus four thicknesses 
of the MgO barrier for PC, and the data is in perfect consistency with the 
physics of tunneling.

In summary, we have analyzed non-equilibrium quantum transport properties of 
Fe/MgO/Fe MTJs from atomic first principles without any phenomenological 
parameter. All the obtained voltage scales for transport features are 
consistent with experimental data, these include the turning on voltage for 
currents, voltage scale for TMR quenching, maximum value of $V_{out}$ as well 
as the turning over voltage of $V_{out}$. The quench of TMR by bias is found 
to be due to a relatively fast increase of channel currents in APC. Very large 
TMR at zero bias is obtained which is stable against small changes 
of atomic positions. The zero bias TMR is, however, expected to be sensitive 
to presence or absence of impurities, oxidization layers and experimental
processing procedures. It is hopeful that even greater TMR can be obtained 
experimentally.

\noindent
{\bf Acknowledgments:}
We gratefully acknowledge financial support from NSERC of Canada, FRQNT of 
Quebec, CIAR, and Killam Research Fellowship (H.G), and NSF-China (K.X).
We thank Dr. Lei Liu for assistance in several useful analysis software 
tools, Dr. Eric Zhu for his participation at early stages of this work, and
Dr. Xiaoguang Zhang for illuminating communications on TMR physics.


\begin{thebibliography}{00}
\bibitem{foot2}
The bands in Fig.3a,b are at $|k_x|=k_y|=0.12$ where the APC transmission 
have the broad peaks (Fig.3c,d). On the other hand, we refer bands at 
$k_x=k_y=0$ as $\Delta_1$ band etc.. In this sense, band-1 in Fig.3a is rather 
like the $\Delta_1$ band while band-2 of Fig.3b is like the $\Delta_5$ band. 
Coupling between these bands in APC are possible due $|k_x|=k_y|\neq 0$ and 
due to scattering at the interfaces.
\end{thebibliography}
\end{document}